\documentclass[aps,prd,twocolumn,showpacs,floatfix,preprintnumbers,amsmath,amssymb,nofootinbib,groupedaddress, superscriptaddress]{revtex4-1}

\input epsf
\usepackage{graphicx}
\usepackage{color}
\usepackage{mathtools}
\usepackage{mathbbol}

\newcommand{\beq}{\begin{equation}}
\newcommand{\eeq}{\end{equation}}
\newcommand{\barr}{\begin{eqnarray}}
\newcommand{\earr}{\end{eqnarray}}

\newcommand{\bs}{\boldsymbol}

\newcommand{\apjl}{Astrophys. J. Lett.}
\newcommand{\aap}{Astron. Astrophys.}

\newcommand{\mnras}{Mon. Not. R. Astron. Soc.}

\newcommand{\physrep}{Phys. Rep.}
\newcommand{\araa}{Annual Rev. Astron. Astrophys.}
\newcommand{\jcap}{J. Cosm. Astropart. Phys.}
\newcommand{\apss}{Astrophys. and Space Science}

\usepackage{color}
\newcommand{\changeJ}[1]{{\textcolor{black}{#1}}}
\newcommand{\newversion}[1]{{\textcolor{black}{#1}}}

\newcommand{\lsim}{\mathrel{\hbox{\rlap{\lower.55ex\hbox{$\sim$}} \kern-.3em \raise.4ex \hbox{$<$}}}}
\newcommand{\gsim}{\mathrel{\hbox{\rlap{\lower.55ex\hbox{$\sim$}} \kern-.3em \raise.4ex \hbox{$>$}}}}

\begin{document}
\title{Constraints on dark matter interactions with standard model particles from CMB spectral distortions}
\author{Yacine Ali-Ha\"imoud}
\affiliation{Department of Physics and
  Astronomy, Johns Hopkins University, 3400 N.~Charles Street, Baltimore, MD 21218}
\author{Jens Chluba}
\affiliation{Department of Physics and
  Astronomy, Johns Hopkins University, 3400 N.~Charles Street, Baltimore, MD 21218}
\affiliation{Kavli Institute for Cosmology Cambridge, Madingley Road, Cambridge, CB3 0HA, U.K}
\author{Marc Kamionkowski}
  \affiliation{Department of Physics and
  Astronomy, Johns Hopkins University, 3400 N.~Charles Street, Baltimore, MD 21218}
\date{\today}

\begin{abstract}

We propose a new method to constrain elastic scattering between dark matter (DM) and standard model particles in the early Universe. Direct or indirect thermal coupling of non-relativistic DM with photons leads to a heat sink for the latter. This results in spectral distortions of the cosmic microwave background (CMB), the amplitude of which can be as large as a few times the DM-to-photon number ratio.  We compute CMB spectral distortions due to DM-proton, DM-electron and DM-photon scattering for generic energy-dependent cross sections and DM mass $m_{\chi} \gtrsim 1$ keV. Using FIRAS measurements we set constraints on the cross sections for $m_\chi \lesssim 0.1$ MeV. In particular, for energy-independent scattering we obtain $\sigma_{\textrm{DM-proton}} \lesssim 10^{-24} \textrm{cm}^2~ (\textrm{keV}/m_{\chi})^{1/2}$, $\sigma_{\textrm{DM-electron}} \lesssim 10^{-27} \textrm{cm}^2~ (\textrm{keV}/m_{\chi})^{1/2}$ and $\sigma_{\textrm{DM-photon}} \lesssim 10^{-39} \textrm{cm}^2~ (m_{\chi}/\textrm{keV})$. An experiment with the characteristics of PIXIE would extend the regime of sensitivity up to masses $m_\chi \sim 1$ GeV. 

\end{abstract}

\maketitle

\textbf{Introduction} -- In the standard cosmological model, dark matter (DM) only interacts gravitationally, and a single number (its abundance) suffices to accurately describe its effect on the cosmic microwave background (CMB) anisotropies and large-scale structure (LSS). Yet, a variety of DM candidates are predicted to interact weakly with standard model particles \cite{Jungman_1996, Feng_2010}. Several direct detection experiments \cite{Behnke_2011, Xenon100_2012, Ahmed_2011, Essig_2012}, and various astronomical and cosmological probes \cite{Feng_2010, Peter_2012} have or will set limits on interacting DM models, but large portions of parameter space are still unconstrained.

In this \changeJ{Letter,} we propose a new probe of DM interactions with standard model particles, relying on spectral distortions \newversion{(hereafter, SDs)} of the CMB blackbody spectrum. It is well known that energy injection \changeJ{at redshifts} $z \lesssim 2 \times 10^6$ can distort the CMB spectrum \changeJ{\cite{Zeldovich1969, Sunyaev1970mu, Hu_1993}}. \newversion{SDs take the form of a chemical-potential (``$\mu$-type") distortion if energy injection occurs early enough ($z \gtrsim 5 \times 10^4$) that Compton scattering efficiently redistributes photons in frequency. The SD shape smoothly transitions to a ``$y$-type" distortion at lower redshifts when photon frequency redistribution becomes less efficient. SDs have} been used, for example, to constrain decaying \changeJ{particles \cite{Hu_1993b, Chluba2013fore} or the primordial small-scale power spectrum \citep{Hu1994, Chluba2012inflaton}}. Conversely, any energy \emph{extraction} from the photons can also induce SDs. Such a process is at work in the standard cosmological scenario \changeJ{\cite{Chluba_2012, Khatri_2012, Pajer_2013}:} CMB photons heat up the non-relativistic plasma through \changeJ{Compton} scattering followed by rapid Coulomb interactions. Instead of cooling down adiabatically with a temperature $T_b \propto 1/a^2$, where $a$ is the scale factor, electrons, protons and helium nuclei (``baryons" for short) are maintained in thermal equilibrium with the photons down to redshift $z \approx 200$, with $T_b \approx T_\gamma \propto 1/a$ \cite{Peebles_1968}. This process removes heat from the CMB and leads to \emph{negative} $\mu$-type and $y$-type distortions of a few times the baryon-to-photon number ratio, of order a few parts in a billion. 

The effect we describe in this work is an extension of the aforementioned process. If a non-relativistic DM particle is thermally coupled to the electron-nucleon plasma through frequent collisions, the energy extracted from the photons by Compton scattering is then redistributed among a larger number of particles.
Maintaining the baryons \emph{and} the DM in thermal equilibrium with the radiation therefore requires an increased rate of energy extraction from the photons. Direct scattering of DM particles with photons have the same outcome. In both cases SDs are generated, of order up to a few times the DM-to-photon number ratio. Given the known DM mass density, this number is inversely proportional to the DM mass. 

To be more precise, this effect only takes place as long as the heating of the DM is more efficient than the competing adiabatic cooling due to cosmological expansion. Hence the final SD is \emph{at most} a few times the DM-to-photon number ratio. It is smaller the shorter the epoch of tight thermal coupling of DM with the photon-baryon plasma is. For a given sensitivity to SDs, there is therefore a maximum mass that can be probed, such that the SD reaches the instrument sensitivity when the DM is tightly coupled at all relevant times. This maximum mass is $\sim 0.1$ MeV for FIRAS \cite{Fixen_1996} and will be $\sim 1$ GeV for a future SD experiment with the sensitivity of PIXIE \cite{Kogut_2011}. Below that maximal mass, SD measurements can set an upper bound to the duration of tight coupling, hence to the scattering cross section. 

Since the maximal SD is inversely proportional to the DM mass, the effect we introduce allows to test light-DM models, while most direct-detection experiments are insensitive to sub-GeV masses. In this first study we limit ourselves to DM masses greater than a keV, such that the DM is already non-relativistic at the beginning of the distortion epoch, $z \approx 2 \times 10^{6}$.

\textbf{DM scattering with protons or electrons} -- We assume that the DM particle $\chi$ can elastically scatter off baryons with a momentum-transfer cross section of the form $\sigma(v) = \sigma_n v^n$, 
where $n$ is an arbitrary integer and $v$ is the magnitude of the baryon-DM relative velocity (in units of the speed of light). \newversion{Some of the best-motivated DM models indeed have interactions of this form \cite{Gluscevic_2015}. For instance, electric-dipole or magnetic-dipole DM coupling to the standard model through heavy charged messengers have cross sections $\sigma \propto v^2$ and $v^4$, respectively. In this Letter we} specifically consider scattering with either protons or free electrons, and neglect scattering with helium nuclei.

Baryons maintain a single temperature at all times due to extremely frequent Coulomb scatterings \cite{Peebles_1968}. The baryon temperature $T_b$ is in turn closely coupled to the CMB temperature $T_{\gamma}$ through Compton scattering of photons by free electrons. As long as the rate of DM-baryon scattering is much larger than the Hubble expansion rate, DM particles have a Maxwellian velocity distribution with temperature $T_{\chi} \approx T_b$. Once the collision rate falls below the expansion rate, then $(i)$ the DM momenta start redshifting freely as $p_{\chi} \propto (1+z)$, and $(ii)$ if this decoupling is not instantaneous and is velocity-dependent, the DM velocity distribution $f_{\chi}(\bs{v}, t)$ is no longer necessarily described by a Maxwell-Boltzmann law. One should then in principle compute $f_{\chi}$ by solving \changeJ{the Boltzmann} equation.

In the limit that the DM distribution is \changeJ{thermal and} that random velocities dominate over bulk flows (valid for $z \gtrsim 10^4$), Ref.~\cite{Dvorkin_2014} have showed that the evolution of the DM temperature is governed by
\barr
\dot{T}_{\chi} &=& - 2 H T_{\chi} + \Gamma_{\chi b} \left(T_b - T_{\chi}\right),\label{eq:Tchi} \\
\textrm{\changeJ{with}} \ \ \ \ \Gamma_{\chi b} &\equiv& \frac{2 c_n N_b \sigma_n m_b m_{\chi}}{(m_b + m_{\chi})^2}\left(\frac{T_b}{m_b} + \frac{T_{\chi}}{m_{\chi}} \right)^{\frac{n+1}2}, \label{eq:Gamma_cb}
\earr
where $c_n$ is a constant depending on $n$ given in \cite{Dvorkin_2014} and $N_b = N_b^0 a^{-3}$ is the number density of scattering \changeJ{baryons.} Assuming radiation domination (valid at $z \gtrsim 3000$), the Hubble expansion rate $H(a)$ scales as $H = H_0 (\Omega_r^0)^{1/2} a^{-2}$. Setting $T_b = T_\gamma = T_{\gamma}^0 a^{-1}$, the ratio $\Gamma_{\chi b}/H$ takes the form
\beq
\frac{\Gamma_{\chi b}}{H} = \left(\frac{a_{\chi b}}{a}\right)^{\frac{n+3}{2}} \left(\frac{m_\chi/ m_b + T_\chi/T_b}{m_\chi/m_b + 1}\right)^{\frac{n+1}{2}},\label{eq:Gamma-H-ratio}
\eeq
where we have defined the characteristic scale factor $a_{\chi b}$ such that: 
\barr
(a_{\chi b})^{\frac{n+3}2} \equiv \frac{m_b}{m_\chi} \left(1 + \frac{m_b}{m_\chi}\right)^{\frac{n-3}{2}}\frac{2 c_n \sigma_n N_b^0 \left(\frac{T_\gamma^0}{m_b}\right)^\frac{n+1}2}{H_0 (\Omega_r^0)^{1/2}}.\label{eq:achib}
\earr
The scale factor $a_{\chi b}$ marks the transition between thermal coupling $(\Gamma_{\chi b} \gtrsim H)$ and decoupling $(\Gamma_{\chi b} \lesssim H$). In this \changeJ{work,} we will only consider slopes $n > -3$ for which the DM starts tightly coupled to the baryons at early times, and thermally decouples at $a \gtrsim a_{\chi b}$. In these cases DM-baryon scattering can have negligible effects on CMB anisotropies and LSS, yet can still manifest itself at high \changeJ{redshifts} and induce SDs. 

We define the dimensionless parameter 
\beq
r_{\chi b} \equiv \frac{\Gamma_{\chi b}(T_b - T_\chi)}{H T_b}, \label{eq:r(a)}
\eeq
characterizing the efficiency of the DM-baryon thermal coupling. In the limit that $\Gamma_{\chi b} \gg H$, DM and baryon temperatures are tightly coupled, $T_\chi \approx T_b = T_{\gamma}$\changeJ{, which} implies $\dot{T}_{\chi} \approx - H T_\gamma$. Inserting this value \changeJ{back into the left-hand-side of} Eq.~\eqref{eq:Tchi} we obtain $r_{\chi b}(a \ll a_{\chi b}) = 1$. On the other hand, $r_{\chi b}(a \gg a_{\chi b}) \rightarrow 0$ since $\Gamma_{\chi b} \ll H$ in that regime. The transition from tight coupling to decoupling is sharper as $n$ is increased, as can be seen from Eq.~\eqref{eq:Gamma-H-ratio}. For $n = -2$, $\Gamma_{\chi b}/H \sim \sqrt{a_{\chi b}/a}$ and thermal decoupling is very gradual, taking a few decades in scale factor. This implies that the description of the DM velocity distribution with a Maxwell-Boltzmann law is very inaccurate in that case (unless the DM is self-interacting, which we do not consider here). We defer the full solution of the collisional Boltzmann equation to future work, and will focus on $n \geq -1$ in what follows. 

We confirm all these features by solving Eq.~\eqref{eq:Tchi} numerically and computing the resulting $r_{\chi b}(a)$. We find that $r_{\chi b} \approx 1/2$ at $a \approx (2/3) a_{\chi b}$ for a broad range of mass ratios and slopes $n \geq -1$. Once the DM thermally decouples from the baryons, its velocity distribution is no longer Maxwell-Boltzmann and Eq.~\eqref{eq:Tchi} \changeJ{is} no longer valid. As a result the parameter $r_{\chi b}(a)$ we computed is inaccurate for $a \gtrsim a_{\chi b}$. We therefore adopt the following simple step-function approximation for $r_{\chi b}(a)$, which captures its essential features:
\beq
r_{\chi b}(a) = 1 \  \textrm{ for } \ a \leq (2/3) a_{\chi b}  \ \textrm{ and  }  0  \textrm{ otherwise}.
\eeq

\textbf{DM-photon scattering} -- We also consider direct scattering of DM particles with photons, with an energy-dependent momentum-transfer cross-section $\sigma(E_\gamma) = \sigma_p (E_{\gamma}/E_0)^p$. We set the normalization at $E_0 = 1$ keV, close to the characteristic CMB photon energy at $z \sim 10^6$. A Thomson-like scattering ($p=0$) would arise, e.g., for a millicharged DM. A quadratic dependence on energy ($p = 2$) could arise from a DM particle with an electric or magnetic dipole moment \cite{Sigurdson_2004}. A quartic dependence ($p=4$) occurs e.g.~for Rayleigh DM \cite{Weiner_2012}. Assuming the DM is non-relativistic ($m_{\chi} \gg T_{\gamma}$), the resulting heating rate is easily obtained by generalizing the calculation of the Compton heating rate due to scattering off free electrons \cite{Weymann_1965}: 
\barr
\dot{T}_\chi|_{\chi \gamma} &=& \Gamma_{\chi \gamma} (T_\gamma - T_\chi), \\
 \textrm{with} \ \ \Gamma_{\chi \gamma} &\equiv& \frac83 \frac{d_p \sigma_p (T_{\gamma}/E_0)^p \rho_\gamma}{m_\chi},
\earr
where $d_p$ is a numerical constant, with value (0.28, 1, 4.8, 28.2, 1558) for $p = (-1, 0, 1, 2, 4)$ respectively.

In this \changeJ{case,} the ratio of interaction to expansion rates takes the form
\barr
\frac{\Gamma_{\chi \gamma}}{H} &=& \left(\frac{a_{\chi \gamma}}{a}\right)^{p+2}, \\
\textrm{where} \ \ (a_{\chi \gamma})^{p+2} &\equiv& \frac83 \frac{d_p \sigma_p \rho_\gamma^0}{m_\chi H_0 (\Omega_r^0)^{1/2}} (T_{\gamma}^0/E_0)^p. \label{eq:achig}
\earr
We only consider $p > -2$ so that DM-photon interactions are efficient at early times and irrelevant at $a \gtrsim a_{\chi \gamma}$. Here again we define the dimensionless parameter $r_{\chi \gamma} \equiv \Gamma_{\chi \gamma}(T_{\gamma} - T_{\chi})/H T_{\gamma}$, and compute it numerically (assuming DM scatters off photons only). Its behavior is very similar to the one obtained for $r_{\chi b}$, with the correspondence $n \leftrightarrow 2 p + 1$, and we also approximate it by a step function \changeJ{with a transition at $a = (2/3) a_{\chi \gamma}$}.

\textbf{Spectral distortions} -- The baryon temperature evolves according to $(i)$ adiabatic cooling due to cosmological expansion, $(ii)$ Compton heating by CMB photons and $(iii)$ energy exchange with the DM: 
\beq
\dot{T}_b = - 2 H T_b + \Gamma_{\rm Com}(T_{\gamma} - T_b) + \frac{N_\chi}{N_{b}^{\rm tot}} \Gamma_{\chi b}(T_{\chi} - T_b). \label{eq:dot-Tb}
\eeq
Here $N_\chi \equiv \rho_\chi/m_\chi$ is the number density of DM particles and $N_{b}^{\rm tot}$ is the total number density of ``baryons" (nuclei and free electrons) maintained in equilibrium with one another. As long as the plasma is fully ionized (valid for $z \gtrsim 6000$), $N_b^{\rm tot}=  2 N_{\rm H} + 3 N_{\rm He} \approx  \rho_b/m_{\rm H}(2 - \frac54 Y_{\rm He})$ where \newversion{$N_{\rm H}$ and $N_{\rm He}$ are the abundances of hydrogen and helium and} $Y_{\rm He}$ is the helium fraction by mass. The last term in Eq.~\eqref{eq:dot-Tb} is easily obtained from Eq.~\eqref{eq:Tchi} by requiring that baryon-DM collisions conserve the total thermal energy.

The rate of extraction of energy from the photons by Compton scattering is \cite{Chluba_2012}
\beq
\rho_\gamma \frac{d}{dt}\left( \frac{\Delta \rho_\gamma}{\rho_\gamma}\right)_{\rm Com}  = \frac32 N_{b}^{\rm tot} \Gamma_{\rm Com}(T_b - T_{\gamma}). \label{eq:drho_gamma}
\eeq
Since for all redshifts $z \gtrsim 200$ Compton scattering maintains $T_b \approx T_{\gamma}$ to very high accuracy, the rate of change of the baryon temperature is just $\dot{T}_b = - H T_b$. Replacing the left-hand-side of Eq.~\eqref{eq:dot-Tb} by this value we obtain the net Compton heating rate, and substitute it in Eq.~\eqref{eq:drho_gamma} to arrive at
\beq
\rho_\gamma \frac{d}{dt}\left( \frac{\Delta \rho_\gamma}{\rho_\gamma}\right)_{\rm Com} = - \frac32 (N_b^{\rm tot} + r_{\chi b} N_\chi) H T_\gamma, \label{eq:drho_gamma-final}
\eeq
where the parameter $r_{\chi b}$ was defined in Eq.~\eqref{eq:r(a)}.
The first term was derived in Ref.~\cite{Chluba_2012} and translates the extraction of energy from photons due to Compton heating of the baryons. The second term arises from the enhanced heat capacity of the baryon-DM fluid due to DM-baryon scattering.


The effect of direct DM-photon scattering is similar: in this case the cooling rate of the photons is given by
\barr
\rho_\gamma \frac{d}{dt}\left( \frac{\Delta \rho_\gamma}{\rho_\gamma}\right)_{\chi \gamma} = - \frac32 r_{\chi \gamma}  N_\chi H T_\gamma.  \label{eq:dist_gamma_chi}
\earr

To obtain the final relative amplitude of \changeJ{spectral} distortions, we have to integrate Eq.~\eqref{eq:drho_gamma-final} or \eqref{eq:dist_gamma_chi} over time. The high-redshift boundary is at \changeJ{$z_{\mu} \approx 2 \times 10^{6}$}: energy injection (or extraction) at $z \gtrsim z_{\mu}$ simply leads to a change of the photon temperature and does not distort the blackbody spectrum \cite{Hu_1993}. The low-redshift end is in principle $z_{\rm min} \approx 200$, the epoch of thermal decoupling of baryons from photons. In practice, energy injection at $z \lesssim 10^4$ leads mostly to a $y$-type distortion \cite{Hu_1993, Chluba_2013}. Compton scattering by free electrons in hot clusters and the reionized intergalactic medium leads to a $y$-distortion of $\sim 2 \times 10^{-6}$ \cite{Hill_2015}. This is below the sensitivity of FIRAS \cite{Fixen_1996}, but orders of magnitude larger than that of PIXIE \cite{Kogut_2011}. We shall therefore not consider pure $y$-distortions in this work, and cut the integration at $z_{\rm min} = 10^4$ ($a_{\max} = 10^{-4}$). This allows us to consider a radiation-dominated and fully ionized universe, and neglect bulk flows relative to thermal velocities \cite{Dvorkin_2014}.


%

Our final estimate for the amplitude of SDs due to DM-baryon collisions is therefore
\barr
\Delta &\equiv& \frac{\Delta \rho_\gamma}{\rho_\gamma} \approx- \frac32 \int_{t(z_{\mu})}^{t(z_{\rm min})}  \frac{N_b^{\rm tot} + r_{\chi b} ~ N_{\chi}}{\rho_\gamma} \changeJ{\,T_\gamma \, H} d t\nonumber\\
& \approx& -0.56  \left[ \frac{N_b^{\rm tot}}{N_\gamma} \log\left(\frac{a_{\rm max}}{a_{\mu}}\right) +   \frac{N_\chi}{N_\gamma} \log\left(\frac{a_*}{a_{\mu}}\right)  \right],~\label{eq:final}
\earr
where \changeJ{$N_\gamma \approx \rho_\gamma/(2.7 \,T_\gamma)$} is the number density of CMB photons, and we have taken the constant baryon-to-photon and DM-to-photon number ratios out of the integrals. The cutoff $a_*$ is defined as \changeJ{$a_* \equiv \max[ a_{\mu}, \min(a_{\rm max}, (2/3)a_{\chi b})]$} (we smooth the transitions for better visual results). The same expression applies to DM-photon scattering with the substitution $a_{\chi b} \rightarrow a_{\chi \gamma}$. The integrals only depend logarithmically on the \changeJ{boundaries.} Our various approximations (\changeJ{i.e.,} taking a sharp boundary at $a_{\mu} = z^{-1}_{\mu}$ instead of using a SD visibility function \cite{Chluba_2012}, assuming a step function for $r_{\chi b}(a)$ and choosing $a_{\rm max} = 10^{-4}$) should therefore not significantly affect our results. They have the advantage of giving simple analytic expressions. 

\begin{figure}
\includegraphics[width = 86 mm]{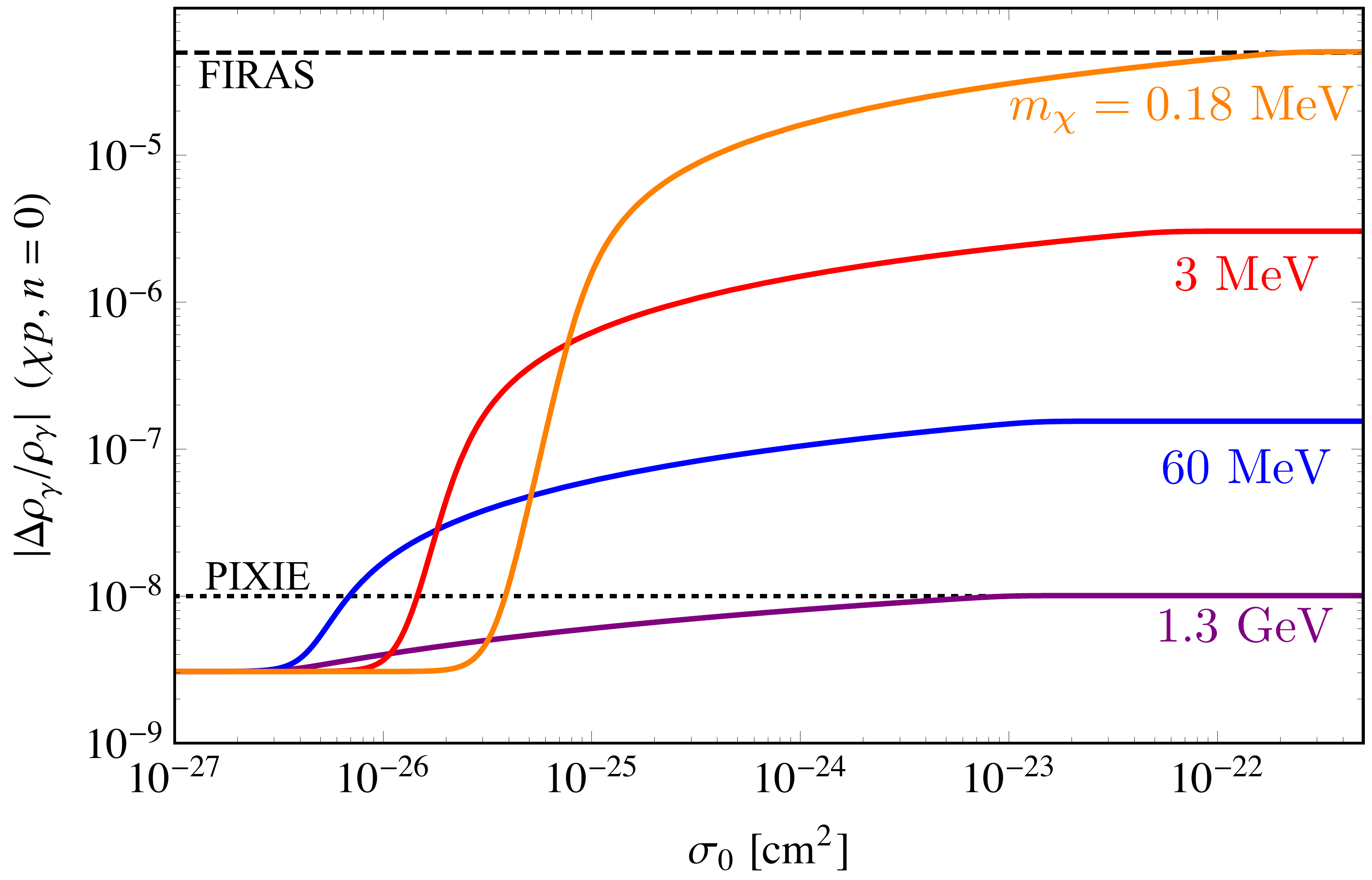}
\caption{Absolute value of the photon distortion $\Delta \rho_\gamma/\rho_\gamma$ for DM collisions with \emph{protons}, for a velocity-independent cross section $\sigma_0$. The solid curves are labelled by the DM particle mass. The upper dashed curve indicates the approximate constraint from FIRAS $\Delta \rho_{\gamma}/\rho_{\gamma} \leq 5 \times 10^{-5}$ \cite{Fixen_1996}. The lower dotted curve indicates the approximate forecasted sensitivity of PIXIE $\Delta \rho_\gamma/\rho_\gamma \sim 10^{-8}$ \cite{Kogut_2011}. }\label{fig:dist}
\end{figure}

\textbf{Results} -- All our results are computed with the current best-fit values for cosmological parameters \cite{Fixen_1996, Planck_2015}. We show in Fig.~\ref{fig:dist} the photon distortion $\Delta \rho_\gamma/\rho_\gamma$ for velocity-independent DM-proton scattering as a function of $\sigma_0$, for several values of the DM mass. For cross sections small enough that $(2/3) a_{\chi b} < a_{\mu}$ ($a_* = a_{\mu}$), DM scattering has no effect and the distortion plateaus at $\Delta_0 \approx- 3 \times 10^{-9}$, due exclusively to the cooling of baryons \cite{Chluba_2012, Khatri_2012, Pajer_2013}. For cross sections large enough that $(2/3)a_{\chi b} > a_{\max}$, the DM is tightly coupled to baryons at all relevant times, and the distortion saturates ($a_* = a_{\max}$ regardless of the cross section). In this limit, the distortion is enhanced by a factor $N_{\chi}/N_b^{\rm tot} \approx 3 (\textrm{GeV}/m_\chi)$, so that $\Delta [\sigma_n \rightarrow \infty] \approx \Delta_0 -9 \times 10^{-6} (\textrm{MeV}/m_\chi)$. Between these two regimes, the distortion scales logarithmically with $a_{\chi b}$ \changeJ{and} hence $\sigma_n$.

For a given sensitivity $\Delta_{\max}$, a maximum mass $m_{\chi}^{\max}$ can be probed, such that $|\Delta[\sigma_n \rightarrow \infty, m_{\chi}^{\max}]| = \Delta_{\max}$:  
\beq
m_{\chi}^{\max} \approx 1 ~ \textrm{MeV} \times \frac{9 \times 10^{-6}}{\Delta_{\max} - 3 \times 10^{-9}}. \label{eq:mmax}
\eeq
Higher masses are completely unconstrained as they lead to a too small number density of DM particles. FIRAS has measured the CMB blackbody spectrum to an accuracy $\Delta_{\max} \approx 5 \times 10^{-5}$ \cite{Fixen_1996}, implying $m_{\chi}^{\max} \approx 0.18$ MeV. The proposed experiment PIXIE \cite{Kogut_2011} will reach a sensitivity $\Delta_{\max} \approx 10^{-8}$. This will allow to constrain the DM-baryon and DM-photon cross sections up to DM masses $m_{\chi}^{\max} \approx 1.3$ GeV.

For masses $m_{\chi} \leq m_{\chi}^{\max}$, a measurement of the CMB blackbody spectrum to a precision $\Delta_{\max}$ would imply an upper limit on the cross sections $\sigma_n^{\chi b}$ or $\sigma_p^{\chi \gamma}$. For DM-baryon collisions we obtain, using Eqs.~\eqref{eq:final} and \eqref{eq:achib}, 
\barr
\sigma_n^{\chi b} &\leq& C_n  \frac{m_\chi}{m_b}\left(1 + \frac{m_b}{m_\chi}\right)^{\frac{3-n}2} \left( \frac{a_{\max}}{a_{\mu}}\right)^{\frac{n+3}{2} m_{\chi}/m_{\chi}^{\max}}. ~~\label{eq:sigma_chib_max}
\earr
For DM-proton collisions, the numerical constants $C_n$ are $(1.4 \times 10^{-30}, 1.1 \times 10^{-27}, 8.2 \times 10^{-25}, 5.5 \times 10^{-22})$ cm$^2$ for $n = (-1, 0, 1, 2)$ respectively. For DM-electron collisions, the corresponding values are $(1.4 \times 10^{-30}, 2.6 \times 10^{-29}, 4.5 \times 10^{-28}, 7.0 \times 10^{-27})$ cm$^2$. The constraint on the DM-photon cross section is obtained similarly from Eqs.~\eqref{eq:final} and \eqref{eq:achig}: 
\beq
\sigma_p^{\chi \gamma} \lesssim D_p \frac{m_{\chi}}{\textrm{MeV}} \left( \frac{a_{\max}}{a_{\mu}}\right)^{(p+2) m_{\chi}/m_{\chi}^{\max}}, \label{eq:sigma_chig_max}
\eeq
with $D_p = ( 6.3, ~5.6, ~3.7, ~2.0, ~0.4) \times 10^{-37}$ cm$^2$ for $p = (-1, 0, 1, 2, 4)$, respectively. 

Equations \eqref{eq:mmax}, \eqref{eq:sigma_chib_max} and \eqref{eq:sigma_chig_max} are the main results of this Letter. Given a sensitivity $\Delta_{\max}$, they allow to obtain upper limits on DM-baryon and DM-photon cross sections with power-law dependence on the baryon-DM relative velocity or photon energy \newversion{(with $n, p \geq -1$)}, up to a maximal DM mass $m_{\chi}^{\max}$. 

\begin{figure}
\includegraphics[width = 86 mm]{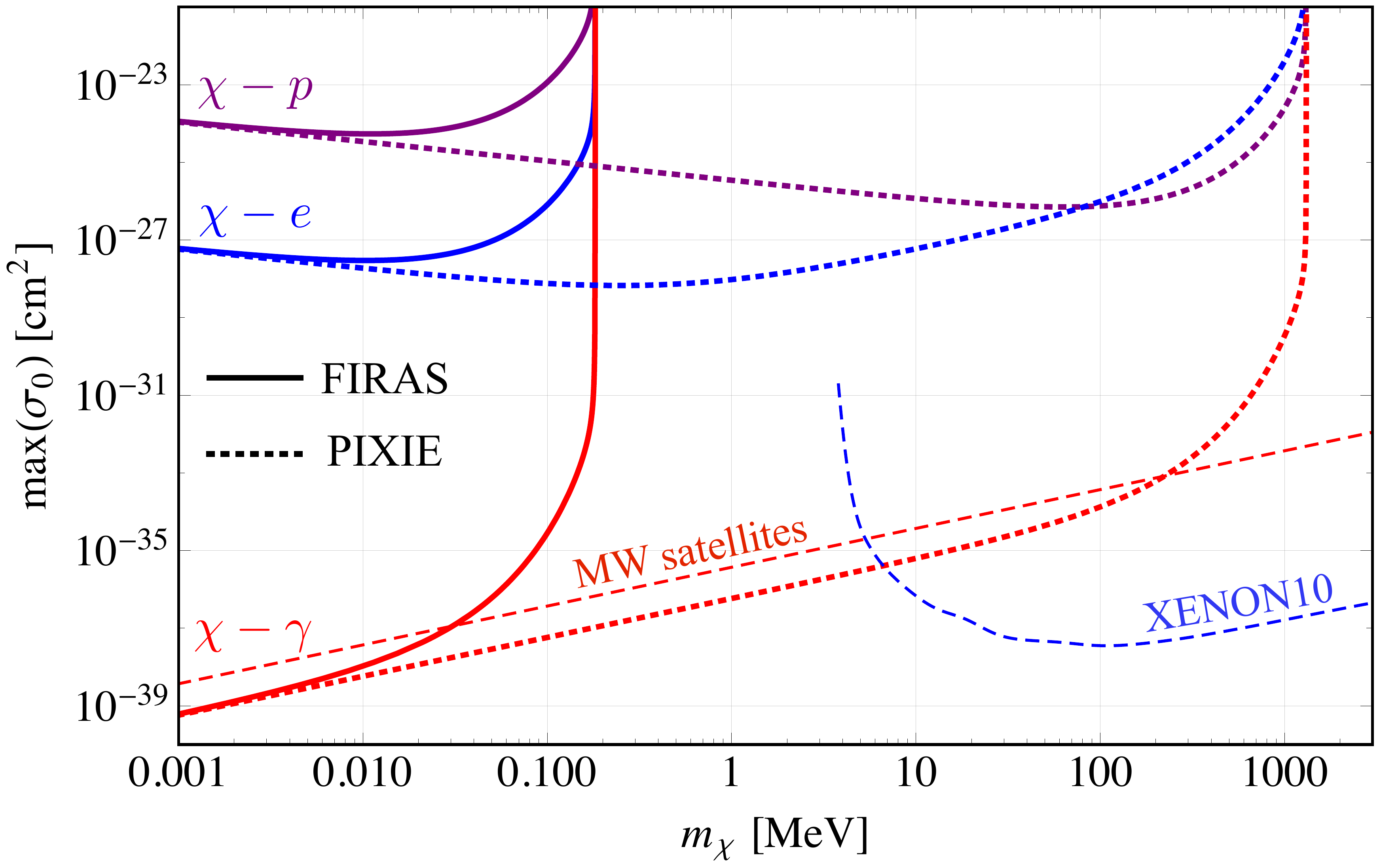}
\caption{Current upper bounds from FIRAS (solid) and forecasted detection thresholds from PIXIE (dotted) on the energy-independent DM-proton (purple), DM-electron (blue) and DM-photon (red) cross sections $\sigma_0$, as a function of the DM mass. Masses $m_\chi \geq 0.18$ MeV are unconstrained by FIRAS as the distortion can never reach $\Delta \rho_\gamma/\rho_\gamma = 5 \times 10^{-5}$, even for infinititely large cross section. PIXIE will extend the domain of constrainable masses by four orders of magnitude, up to $m_\chi \approx 1.3$ GeV. For comparison, we also show the constraints on DM-electron scattering from XENON10 data \cite{Essig_2012} and the limits on DM-photon scattering from Milky Way satellite counts \cite{Boehm_2014}. No other probe currently constrains DM-proton scattering in the range of masses shown.}\label{fig:constraints}
\end{figure}

We plot in Fig.~\ref{fig:constraints} the current constraints on the energy-independent cross sections $\sigma_0^{\chi p}, \sigma_0^{\chi e}, \sigma_0^{\chi \gamma}$ as a function of the DM mass given the FIRAS measurements. We also show the forecasted constraints for the sensitivity of PIXIE.

%

\textbf{Comparison with previous bounds} -- Most direct detection experiments only constrain DM-nucleon cross sections for masses $m_{\chi} \gtrsim$ few GeV, required to produce sufficient nuclear recoil. Ref.~\cite{Dvorkin_2014} derive constraints on the ratio $\sigma_n/m_\chi$ for DM-proton collisions in the limit $m_\chi \gg m_{\rm H}$, using CMB anisotropy and LSS data. SDs therefore provide a probe of DM-nuclei scattering in a mass range complementary to the one currently constrained. In particular, our limits on DM-proton scattering from FIRAS measurements are the only existing bounds for $m_\chi \lesssim 0.1$ MeV.

Ref.~\cite{Essig_2012} have set the first constraints on the scattering of sub-GeV DM with \emph{electrons}, which could lead to ionization events in the target material \cite{Essig_2012a}. For a velocity-independent cross section, they find $\sigma_0 \lesssim 3 \times 10^{-38}$ cm$^2$ for $m_{\chi} = 100$ MeV, significantly better than what we forecast at the same mass for a PIXIE-type experiment, $\sigma_0 \lesssim 10^{-26}$ cm$^2$. The bound of \changeJ{Ref.~\cite{Essig_2012}, however,} worsens rapidly for DM masses below a few MeV. Here again, FIRAS limits give the only existing bounds on DM-electron cross sections for $m_\chi \lesssim 0.1$ MeV.

Ref.~\cite{Boehm_2014} give a constraint on the DM-photon energy-independent cross section using counts of Milky Way satellites, translating to $\sigma_0 \lesssim 3.7 \times 10^{-36} (m_{\chi}/\textrm{MeV})$ cm$^2$. The constraint we set with FIRAS for $m_{\chi} \ll 0.1$ MeV is tighter by a factor of $\sim 5$, and PIXIE will allow to extend it up to $m_\chi \approx 1$ GeV. We also constrain the $p=2$ cross section $\sigma_2 \lesssim 2 \times 10^{-37} (m_\chi/\textrm{MeV})$, tighter by six orders of magnitude than the limit of Ref.~\cite{Wilkinson_2014} using CMB anisotropies.

\textbf{Conclusions} -- We have set forth a new avenue to probe DM interactions with standard model particles, using CMB SDs. We have studied the effect of DM scattering with either protons, electrons or photons, for a power-law velocity and energy dependence of the cross section.  We have shown that the FIRAS measurements can already set constraints on the cross sections for DM masses $m_{\chi} \lesssim 0.1$ MeV. \changeJ{Above this mass, the number of DM particles is too small to affect the effective heat capacity of the plasma at a sufficient level.} The high sensitivity of PIXIE will allow to constrain DM particles with masses up to $\sim 1$ GeV. 

Specific models for the DM particle would predict the shape and relative strengths of interactions with different species. The overall SD can be simply obtained by linearly adding the contributions of each scattering process. While we have only considered the characteristic amplitude of the distortion in this work, its detailed form can easily be obtained by convolving the energy extraction rate with a SD Green's function \cite{Chluba_2013}. In particular, when DM-baryon or DM-photon decoupling occurs after $z \sim 5\times 10^4$, SDs may allow determining specific DM parameters through the residual (non-$\mu$ and non-$y$) distortion \citep{Chluba_2012, Khatri2012mix, Chluba2013PCA}. Delving in such details \changeJ{would, however, require} a better treatment of the DM velocity distribution, which should be computed by solving the Bolzmann equation rather than assuming it is Maxwellian. We have also only considered DM masses $m_{\chi} \gtrsim 1$ keV, such that the DM particle is non-relativistic \changeJ{at} $z \lesssim 2 \times 10^6$. It would be interesting to extend this work to lower masses and relativistic DM particles. We defer the study of these questions to future work.

\small
\textbf{Acknowledgments} -- We thank Daniel Grin, Ely Kovetz and Julian Mu\~noz for useful discussions and comments, and Rouven Essig for sharing the XENON10 constraints. This work was supported at JHU by NSF Grant No.~0244990, NASA NNX15AB18G, the John Templeton Foundation, and the Simons Foundation. \changeJ{JC is supported by the Royal Society as a Royal Society University Research Fellow at the University of Cambridge, U.K.}

\bibliography{dist.bib}

\begin{thebibliography}{33}%
\makeatletter
\providecommand \@ifxundefined [1]{%
 \@ifx{#1\undefined}
}%
\providecommand \@ifnum [1]{%
 \ifnum #1\expandafter \@firstoftwo
 \else \expandafter \@secondoftwo
 \fi
}%
\providecommand \@ifx [1]{%
 \ifx #1\expandafter \@firstoftwo
 \else \expandafter \@secondoftwo
 \fi
}%
\providecommand \natexlab [1]{#1}%
\providecommand \enquote  [1]{``#1''}%
\providecommand \bibnamefont  [1]{#1}%
\providecommand \bibfnamefont [1]{#1}%
\providecommand \citenamefont [1]{#1}%
\providecommand \href@noop [0]{\@secondoftwo}%
\providecommand \href [0]{\begingroup \@sanitize@url \@href}%
\providecommand \@href[1]{\@@startlink{#1}\@@href}%
\providecommand \@@href[1]{\endgroup#1\@@endlink}%
\providecommand \@sanitize@url [0]{\catcode `\\12\catcode `\$12\catcode
  `\&12\catcode `\#12\catcode `\^12\catcode `\_12\catcode `\%12\relax}%
\providecommand \@@startlink[1]{}%
\providecommand \@@endlink[0]{}%
\providecommand \url  [0]{\begingroup\@sanitize@url \@url }%
\providecommand \@url [1]{\endgroup\@href {#1}{\urlprefix }}%
\providecommand \urlprefix  [0]{URL }%
\providecommand \Eprint [0]{\href }%
\@ifxundefined \urlstyle {%
  \providecommand \doi  [0]{\begingroup \@sanitize@url \@doi}%
  \providecommand \@doi [1]{\endgroup \@@startlink {\doibase
  #1}doi:\discretionary {}{}{}#1\@@endlink }%
}{%
  \providecommand \doi  [0]{doi:\discretionary{}{}{}\begingroup
  \urlstyle{rm}\Url }%
}%
\providecommand \doibase [0]{http://dx.doi.org/}%
\providecommand \Doi [0]{\begingroup \@sanitize@url \@Doi }%
\providecommand \@Doi  [1]{\endgroup\@@startlink{\doibase#1}\@@Doi}%
\providecommand \@@Doi [1]{#1\@@endlink}%
\providecommand \selectlanguage [0]{\@gobble}%
\providecommand \bibinfo  [0]{\@secondoftwo}%
\providecommand \bibfield  [0]{\@secondoftwo}%
\providecommand \translation [1]{[#1]}%
\providecommand \BibitemOpen [0]{}%
\providecommand \bibitemStop [0]{}%
\providecommand \bibitemNoStop [0]{.\EOS\space}%
\providecommand \EOS [0]{\spacefactor3000\relax}%
\providecommand \BibitemShut  [1]{\csname bibitem#1\endcsname}%
\bibitem [{\citenamefont {{Jungman}}\ \emph {et~al.}(1996)\citenamefont
  {{Jungman}}, \citenamefont {{Kamionkowski}},\ and\ \citenamefont
  {{Griest}}}]{Jungman_1996}%
  \BibitemOpen
  \bibfield  {author} {\bibinfo {author} {\bibfnamefont {G.}~\bibnamefont
  {{Jungman}}}, \bibinfo {author} {\bibfnamefont {M.}~\bibnamefont
  {{Kamionkowski}}}, \ and\ \bibinfo {author} {\bibfnamefont {K.}~\bibnamefont
  {{Griest}}},\ }\Doi {10.1016/0370-1573(95)00058-5} {\bibfield  {journal}
  {\bibinfo  {journal} {\physrep},\ }\textbf {\bibinfo {volume} {267}},\
  \bibinfo {pages} {195} (\bibinfo {year} {1996})},\ \Eprint
  {http://arxiv.org/abs/hep-ph/9506380} {hep-ph/9506380} \BibitemShut {NoStop}%
\bibitem [{\citenamefont {{Feng}}(2010)}]{Feng_2010}%
  \BibitemOpen
  \bibfield  {author} {\bibinfo {author} {\bibfnamefont {J.~L.}\ \bibnamefont
  {{Feng}}},\ }\Doi {10.1146/annurev-astro-082708-101659} {\bibfield  {journal}
  {\bibinfo  {journal} {\araa},\ }\textbf {\bibinfo {volume} {48}},\ \bibinfo
  {pages} {495} (\bibinfo {year} {2010})},\ \Eprint
  {http://arxiv.org/abs/1003.0904} {arXiv:1003.0904} \BibitemShut {NoStop}%
\bibitem [{\citenamefont {{Behnke}}\ \emph {et~al.}(2011)\citenamefont
  {{Behnke}} \emph {et~al.}}]{Behnke_2011}%
  \BibitemOpen
  \bibfield  {author} {\bibinfo {author} {\bibfnamefont {E.}~\bibnamefont
  {{Behnke}}} \emph {et~al.},\ }\Doi {10.1103/PhysRevLett.106.021303}
  {\bibfield  {journal} {\bibinfo  {journal} {Phys.~Rev.~Lett.},\ }\textbf
  {\bibinfo {volume} {106}},\ \bibinfo {eid} {021303} (\bibinfo {year}
  {2011})},\ \Eprint {http://arxiv.org/abs/1008.3518} {arXiv:1008.3518}
  \BibitemShut {NoStop}%
\bibitem [{\citenamefont {{Aprile}}\ \emph {et~al.}(2012)\citenamefont
  {{Aprile}} \emph {et~al.}}]{Xenon100_2012}%
  \BibitemOpen
  \bibfield  {author} {\bibinfo {author} {\bibfnamefont {E.}~\bibnamefont
  {{Aprile}}} \emph {et~al.},\ }\Doi {10.1103/PhysRevLett.109.181301}
  {\bibfield  {journal} {\bibinfo  {journal} {Phys.~Rev.~Lett.},\ }\textbf
  {\bibinfo {volume} {109}},\ \bibinfo {eid} {181301} (\bibinfo {year}
  {2012})},\ \Eprint {http://arxiv.org/abs/1207.5988} {arXiv:1207.5988}
  \BibitemShut {NoStop}%
\bibitem [{\citenamefont {{Ahmed}}\ \emph {et~al.}(2011)\citenamefont {{Ahmed}}
  \emph {et~al.}}]{Ahmed_2011}%
  \BibitemOpen
  \bibfield  {author} {\bibinfo {author} {\bibfnamefont {Z.}~\bibnamefont
  {{Ahmed}}} \emph {et~al.},\ }\Doi {10.1103/PhysRevD.84.011102} {\bibfield
  {journal} {\bibinfo  {journal} {\prd},\ }\textbf {\bibinfo {volume} {84}},\
  \bibinfo {eid} {011102} (\bibinfo {year} {2011})},\ \Eprint
  {http://arxiv.org/abs/1105.3377} {arXiv:1105.3377} \BibitemShut {NoStop}%
\bibitem [{\citenamefont {{Essig}}\ \emph
  {et~al.}(2012){\natexlab{a}}\citenamefont {{Essig}}, \citenamefont
  {{Manalaysay}}, \citenamefont {{Mardon}}, \citenamefont {{Sorensen}},\ and\
  \citenamefont {{Volansky}}}]{Essig_2012}%
  \BibitemOpen
  \bibfield  {author} {\bibinfo {author} {\bibfnamefont {R.}~\bibnamefont
  {{Essig}}}, \bibinfo {author} {\bibfnamefont {A.}~\bibnamefont
  {{Manalaysay}}}, \bibinfo {author} {\bibfnamefont {J.}~\bibnamefont
  {{Mardon}}}, \bibinfo {author} {\bibfnamefont {P.}~\bibnamefont
  {{Sorensen}}}, \ and\ \bibinfo {author} {\bibfnamefont {T.}~\bibnamefont
  {{Volansky}}},\ }\Doi {10.1103/PhysRevLett.109.021301} {\bibfield  {journal}
  {\bibinfo  {journal} {Phys.~Rev.~Lett.},\ }\textbf {\bibinfo {volume}
  {109}},\ \bibinfo {eid} {021301} (\bibinfo {year} {2012}{\natexlab{a}})},\
  \Eprint {http://arxiv.org/abs/1206.2644} {arXiv:1206.2644} \BibitemShut
  {NoStop}%
\bibitem [{\citenamefont {{Peter}}(2012)}]{Peter_2012}%
  \BibitemOpen
  \bibfield  {author} {\bibinfo {author} {\bibfnamefont {A.~H.~G.}\
  \bibnamefont {{Peter}}},\ }\href@noop {} {\bibfield  {journal} {\bibinfo
  {journal} {ArXiv e-prints}} (\bibinfo {year} {2012})},\ \Eprint
  {http://arxiv.org/abs/1201.3942} {arXiv:1201.3942} \BibitemShut {NoStop}%
\bibitem [{\citenamefont {{Zeldovich}}\ and\ \citenamefont
  {{Sunyaev}}(1969)}]{Zeldovich1969}%
  \BibitemOpen
  \bibfield  {author} {\bibinfo {author} {\bibfnamefont {Y.~B.}\ \bibnamefont
  {{Zeldovich}}}\ and\ \bibinfo {author} {\bibfnamefont {R.~A.}\ \bibnamefont
  {{Sunyaev}}},\ }\Doi {10.1007/BF00661821} {\bibfield  {journal} {\bibinfo
  {journal} {\apss},\ }\textbf {\bibinfo {volume} {4}},\ \bibinfo {pages} {301}
  (\bibinfo {year} {1969})}\BibitemShut {NoStop}%
\bibitem [{\citenamefont {{Sunyaev}}\ and\ \citenamefont
  {{Zeldovich}}(1970)}]{Sunyaev1970mu}%
  \BibitemOpen
  \bibfield  {author} {\bibinfo {author} {\bibfnamefont {R.~A.}\ \bibnamefont
  {{Sunyaev}}}\ and\ \bibinfo {author} {\bibfnamefont {Y.~B.}\ \bibnamefont
  {{Zeldovich}}},\ }\Doi {10.1007/BF00653472} {\bibfield  {journal} {\bibinfo
  {journal} {\apss},\ }\textbf {\bibinfo {volume} {7}},\ \bibinfo {pages} {20}
  (\bibinfo {year} {1970})}\BibitemShut {NoStop}%
\bibitem [{\citenamefont {{Hu}}\ and\ \citenamefont
  {{Silk}}(1993){\natexlab{a}}}]{Hu_1993}%
  \BibitemOpen
  \bibfield  {author} {\bibinfo {author} {\bibfnamefont {W.}~\bibnamefont
  {{Hu}}}\ and\ \bibinfo {author} {\bibfnamefont {J.}~\bibnamefont {{Silk}}},\
  }\Doi {10.1103/PhysRevD.48.485} {\bibfield  {journal} {\bibinfo  {journal}
  {\prd},\ }\textbf {\bibinfo {volume} {48}},\ \bibinfo {pages} {485} (\bibinfo
  {year} {1993}{\natexlab{a}})}\BibitemShut {NoStop}%
\bibitem [{\citenamefont {{Hu}}\ and\ \citenamefont
  {{Silk}}(1993){\natexlab{b}}}]{Hu_1993b}%
  \BibitemOpen
  \bibfield  {author} {\bibinfo {author} {\bibfnamefont {W.}~\bibnamefont
  {{Hu}}}\ and\ \bibinfo {author} {\bibfnamefont {J.}~\bibnamefont {{Silk}}},\
  }\Doi {10.1103/PhysRevLett.70.2661} {\bibfield  {journal} {\bibinfo
  {journal} {Phys.~Rev.~Lett.},\ }\textbf {\bibinfo {volume} {70}},\ \bibinfo
  {pages} {2661} (\bibinfo {year} {1993}{\natexlab{b}})}\BibitemShut {NoStop}%
\bibitem [{\citenamefont {{Chluba}}(2013){\natexlab{a}}}]{Chluba2013fore}%
  \BibitemOpen
  \bibfield  {author} {\bibinfo {author} {\bibfnamefont {J.}~\bibnamefont
  {{Chluba}}},\ }\Doi {10.1093/mnras/stt1733} {\bibfield  {journal} {\bibinfo
  {journal} {\mnras},\ }\textbf {\bibinfo {volume} {436}},\ \bibinfo {pages}
  {2232} (\bibinfo {year} {2013}{\natexlab{a}})},\ \Eprint
  {http://arxiv.org/abs/1304.6121} {arXiv:1304.6121} \BibitemShut {NoStop}%
\bibitem [{\citenamefont {{Hu}}\ \emph {et~al.}(1994)\citenamefont {{Hu}},
  \citenamefont {{Scott}},\ and\ \citenamefont {{Silk}}}]{Hu1994}%
  \BibitemOpen
  \bibfield  {author} {\bibinfo {author} {\bibfnamefont {W.}~\bibnamefont
  {{Hu}}}, \bibinfo {author} {\bibfnamefont {D.}~\bibnamefont {{Scott}}}, \
  and\ \bibinfo {author} {\bibfnamefont {J.}~\bibnamefont {{Silk}}},\ }\Doi
  {10.1086/187424} {\bibfield  {journal} {\bibinfo  {journal} {\apjl},\
  }\textbf {\bibinfo {volume} {430}},\ \bibinfo {pages} {L5} (\bibinfo {year}
  {1994})},\ \Eprint {http://arxiv.org/abs/astro-ph/9402045} {astro-ph/9402045}
  \BibitemShut {NoStop}%
\bibitem [{\citenamefont {{Chluba}}\ \emph {et~al.}(2012)\citenamefont
  {{Chluba}}, \citenamefont {{Erickcek}},\ and\ \citenamefont
  {{Ben-Dayan}}}]{Chluba2012inflaton}%
  \BibitemOpen
  \bibfield  {author} {\bibinfo {author} {\bibfnamefont {J.}~\bibnamefont
  {{Chluba}}}, \bibinfo {author} {\bibfnamefont {A.~L.}\ \bibnamefont
  {{Erickcek}}}, \ and\ \bibinfo {author} {\bibfnamefont {I.}~\bibnamefont
  {{Ben-Dayan}}},\ }\Doi {10.1088/0004-637X/758/2/76} {\bibfield  {journal}
  {\bibinfo  {journal} {\apj},\ }\textbf {\bibinfo {volume} {758}},\ \bibinfo
  {eid} {76} (\bibinfo {year} {2012})},\ \Eprint
  {http://arxiv.org/abs/1203.2681} {arXiv:1203.2681} \BibitemShut {NoStop}%
\bibitem [{\citenamefont {{Chluba}}\ and\ \citenamefont
  {{Sunyaev}}(2012)}]{Chluba_2012}%
  \BibitemOpen
  \bibfield  {author} {\bibinfo {author} {\bibfnamefont {J.}~\bibnamefont
  {{Chluba}}}\ and\ \bibinfo {author} {\bibfnamefont {R.~A.}\ \bibnamefont
  {{Sunyaev}}},\ }\Doi {10.1111/j.1365-2966.2011.19786.x} {\bibfield  {journal}
  {\bibinfo  {journal} {\mnras},\ }\textbf {\bibinfo {volume} {419}},\ \bibinfo
  {pages} {1294} (\bibinfo {year} {2012})},\ \Eprint
  {http://arxiv.org/abs/1109.6552} {arXiv:1109.6552} \BibitemShut {NoStop}%
\bibitem [{\citenamefont {{Khatri}}\ \emph {et~al.}(2012)\citenamefont
  {{Khatri}}, \citenamefont {{Sunyaev}},\ and\ \citenamefont
  {{Chluba}}}]{Khatri_2012}%
  \BibitemOpen
  \bibfield  {author} {\bibinfo {author} {\bibfnamefont {R.}~\bibnamefont
  {{Khatri}}}, \bibinfo {author} {\bibfnamefont {R.~A.}\ \bibnamefont
  {{Sunyaev}}}, \ and\ \bibinfo {author} {\bibfnamefont {J.}~\bibnamefont
  {{Chluba}}},\ }\Doi {10.1051/0004-6361/201118194} {\bibfield  {journal}
  {\bibinfo  {journal} {\aap},\ }\textbf {\bibinfo {volume} {540}},\ \bibinfo
  {eid} {A124} (\bibinfo {year} {2012})},\ \Eprint
  {http://arxiv.org/abs/1110.0475} {arXiv:1110.0475} \BibitemShut {NoStop}%
\bibitem [{\citenamefont {{Pajer}}\ and\ \citenamefont
  {{Zaldarriaga}}(2013)}]{Pajer_2013}%
  \BibitemOpen
  \bibfield  {author} {\bibinfo {author} {\bibfnamefont {E.}~\bibnamefont
  {{Pajer}}}\ and\ \bibinfo {author} {\bibfnamefont {M.}~\bibnamefont
  {{Zaldarriaga}}},\ }\Doi {10.1088/1475-7516/2013/02/036} {\bibfield
  {journal} {\bibinfo  {journal} {\jcap},\ }\textbf {\bibinfo {volume} {2}},\
  \bibinfo {eid} {036} (\bibinfo {year} {2013})},\ \Eprint
  {http://arxiv.org/abs/1206.4479} {arXiv:1206.4479} \BibitemShut {NoStop}%
\bibitem [{\citenamefont {{Peebles}}(1968)}]{Peebles_1968}%
  \BibitemOpen
  \bibfield  {author} {\bibinfo {author} {\bibfnamefont {P.~J.~E.}\
  \bibnamefont {{Peebles}}},\ }\Doi {10.1086/149628} {\bibfield  {journal}
  {\bibinfo  {journal} {\apj},\ }\textbf {\bibinfo {volume} {153}},\ \bibinfo
  {pages} {1} (\bibinfo {year} {1968})}\BibitemShut {NoStop}%
\bibitem [{\citenamefont {{Fixsen}}\ \emph {et~al.}(1996)\citenamefont
  {{Fixsen}} \emph {et~al.}}]{Fixen_1996}%
  \BibitemOpen
  \bibfield  {author} {\bibinfo {author} {\bibfnamefont {D.~J.}\ \bibnamefont
  {{Fixsen}}} \emph {et~al.},\ }\Doi {10.1086/178173} {\bibfield  {journal}
  {\bibinfo  {journal} {\apj},\ }\textbf {\bibinfo {volume} {473}},\ \bibinfo
  {pages} {576} (\bibinfo {year} {1996})},\ \Eprint
  {http://arxiv.org/abs/astro-ph/9605054} {astro-ph/9605054} \BibitemShut
  {NoStop}%
\bibitem [{\citenamefont {{Kogut}}\ \emph {et~al.}(2011)\citenamefont {{Kogut}}
  \emph {et~al.}}]{Kogut_2011}%
  \BibitemOpen
  \bibfield  {author} {\bibinfo {author} {\bibfnamefont {A.}~\bibnamefont
  {{Kogut}}} \emph {et~al.},\ }\Doi {10.1088/1475-7516/2011/07/025} {\bibfield
  {journal} {\bibinfo  {journal} {\jcap},\ }\textbf {\bibinfo {volume} {7}},\
  \bibinfo {eid} {025} (\bibinfo {year} {2011})},\ \Eprint
  {http://arxiv.org/abs/1105.2044} {arXiv:1105.2044} \BibitemShut {NoStop}%
\bibitem [{\citenamefont {{Gluscevic}}\ \emph {et~al.}(2015)\citenamefont
  {{Gluscevic}}, \citenamefont {{Gresham}}, \citenamefont {{McDermott}},
  \citenamefont {{Peter}},\ and\ \citenamefont {{Zurek}}}]{Gluscevic_2015}%
  \BibitemOpen
  \bibfield  {author} {\bibinfo {author} {\bibfnamefont {V.}~\bibnamefont
  {{Gluscevic}}}, \bibinfo {author} {\bibfnamefont {M.~I.}\ \bibnamefont
  {{Gresham}}}, \bibinfo {author} {\bibfnamefont {S.~D.}\ \bibnamefont
  {{McDermott}}}, \bibinfo {author} {\bibfnamefont {A.~H.~G.}\ \bibnamefont
  {{Peter}}}, \ and\ \bibinfo {author} {\bibfnamefont {K.~M.}\ \bibnamefont
  {{Zurek}}},\ }\href@noop {} {\bibfield  {journal} {\bibinfo  {journal} {ArXiv
  e-prints}} (\bibinfo {year} {2015})},\ \Eprint
  {http://arxiv.org/abs/1506.04454} {arXiv:1506.04454} \BibitemShut {NoStop}%
\bibitem [{\citenamefont {{Dvorkin}}\ \emph {et~al.}(2014)\citenamefont
  {{Dvorkin}}, \citenamefont {{Blum}},\ and\ \citenamefont
  {{Kamionkowski}}}]{Dvorkin_2014}%
  \BibitemOpen
  \bibfield  {author} {\bibinfo {author} {\bibfnamefont {C.}~\bibnamefont
  {{Dvorkin}}}, \bibinfo {author} {\bibfnamefont {K.}~\bibnamefont {{Blum}}}, \
  and\ \bibinfo {author} {\bibfnamefont {M.}~\bibnamefont {{Kamionkowski}}},\
  }\Doi {10.1103/PhysRevD.89.023519} {\bibfield  {journal} {\bibinfo  {journal}
  {\prd},\ }\textbf {\bibinfo {volume} {89}},\ \bibinfo {eid} {023519}
  (\bibinfo {year} {2014})},\ \Eprint {http://arxiv.org/abs/1311.2937}
  {arXiv:1311.2937} \BibitemShut {NoStop}%
\bibitem [{\citenamefont {{Sigurdson}}\ \emph {et~al.}(2004)\citenamefont
  {{Sigurdson}}, \citenamefont {{Doran}}, \citenamefont {{Kurylov}},
  \citenamefont {{Caldwell}},\ and\ \citenamefont
  {{Kamionkowski}}}]{Sigurdson_2004}%
  \BibitemOpen
  \bibfield  {author} {\bibinfo {author} {\bibfnamefont {K.}~\bibnamefont
  {{Sigurdson}}}, \bibinfo {author} {\bibfnamefont {M.}~\bibnamefont
  {{Doran}}}, \bibinfo {author} {\bibfnamefont {A.}~\bibnamefont {{Kurylov}}},
  \bibinfo {author} {\bibfnamefont {R.~R.}\ \bibnamefont {{Caldwell}}}, \ and\
  \bibinfo {author} {\bibfnamefont {M.}~\bibnamefont {{Kamionkowski}}},\ }\Doi
  {10.1103/PhysRevD.70.083501} {\bibfield  {journal} {\bibinfo  {journal}
  {\prd},\ }\textbf {\bibinfo {volume} {70}},\ \bibinfo {eid} {083501}
  (\bibinfo {year} {2004})},\ \Eprint {http://arxiv.org/abs/astro-ph/0406355}
  {astro-ph/0406355} \BibitemShut {NoStop}%
\bibitem [{\citenamefont {{Weiner}}\ and\ \citenamefont
  {{Yavin}}(2012)}]{Weiner_2012}%
  \BibitemOpen
  \bibfield  {author} {\bibinfo {author} {\bibfnamefont {N.}~\bibnamefont
  {{Weiner}}}\ and\ \bibinfo {author} {\bibfnamefont {I.}~\bibnamefont
  {{Yavin}}},\ }\Doi {10.1103/PhysRevD.86.075021} {\bibfield  {journal}
  {\bibinfo  {journal} {\prd},\ }\textbf {\bibinfo {volume} {86}},\ \bibinfo
  {eid} {075021} (\bibinfo {year} {2012})},\ \Eprint
  {http://arxiv.org/abs/1206.2910} {arXiv:1206.2910} \BibitemShut {NoStop}%
\bibitem [{\citenamefont {{Weymann}}(1965)}]{Weymann_1965}%
  \BibitemOpen
  \bibfield  {author} {\bibinfo {author} {\bibfnamefont {R.}~\bibnamefont
  {{Weymann}}},\ }\Doi {10.1063/1.1761165} {\bibfield  {journal} {\bibinfo
  {journal} {Physics of Fluids},\ }\textbf {\bibinfo {volume} {8}},\ \bibinfo
  {pages} {2112} (\bibinfo {year} {1965})}\BibitemShut {NoStop}%
\bibitem [{\citenamefont {{Chluba}}(2013){\natexlab{b}}}]{Chluba_2013}%
  \BibitemOpen
  \bibfield  {author} {\bibinfo {author} {\bibfnamefont {J.}~\bibnamefont
  {{Chluba}}},\ }\Doi {10.1093/mnras/stt1025} {\bibfield  {journal} {\bibinfo
  {journal} {\mnras},\ }\textbf {\bibinfo {volume} {434}},\ \bibinfo {pages}
  {352} (\bibinfo {year} {2013}{\natexlab{b}})},\ \Eprint
  {http://arxiv.org/abs/1304.6120} {arXiv:1304.6120} \BibitemShut {NoStop}%
\bibitem [{\citenamefont {{Hill}}\ \emph {et~al.}(2015)\citenamefont {{Hill}},
  \citenamefont {{Battaglia}}, \citenamefont {{Chluba}}, \citenamefont
  {{Ferraro}}, \citenamefont {{Schaan}},\ and\ \citenamefont
  {{Spergel}}}]{Hill_2015}%
  \BibitemOpen
  \bibfield  {author} {\bibinfo {author} {\bibfnamefont {J.~C.}\ \bibnamefont
  {{Hill}}}, \bibinfo {author} {\bibfnamefont {N.}~\bibnamefont {{Battaglia}}},
  \bibinfo {author} {\bibfnamefont {J.}~\bibnamefont {{Chluba}}}, \bibinfo
  {author} {\bibfnamefont {S.}~\bibnamefont {{Ferraro}}}, \bibinfo {author}
  {\bibfnamefont {E.}~\bibnamefont {{Schaan}}}, \ and\ \bibinfo {author}
  {\bibfnamefont {D.~N.}\ \bibnamefont {{Spergel}}},\ }\href@noop {} {\bibfield
   {journal} {\bibinfo  {journal} {ArXiv e-prints}} (\bibinfo {year} {2015})},\
  \Eprint {http://arxiv.org/abs/1507.01583} {arXiv:1507.01583} \BibitemShut
  {NoStop}%
\bibitem [{\citenamefont {{Planck Collaboration}}(2015)}]{Planck_2015}%
  \BibitemOpen
  \bibfield  {author} {\bibinfo {author} {\bibnamefont {{Planck
  Collaboration}}},\ }\href@noop {} {\bibfield  {journal} {\bibinfo  {journal}
  {ArXiv e-prints}} (\bibinfo {year} {2015})},\ \Eprint
  {http://arxiv.org/abs/1502.01589} {arXiv:1502.01589} \BibitemShut {NoStop}%
\bibitem [{\citenamefont {{B{\oe}hm}}\ \emph {et~al.}(2014)\citenamefont
  {{B{\oe}hm}}, \citenamefont {{Schewtschenko}}, \citenamefont {{Wilkinson}},
  \citenamefont {{Baugh}},\ and\ \citenamefont {{Pascoli}}}]{Boehm_2014}%
  \BibitemOpen
  \bibfield  {author} {\bibinfo {author} {\bibfnamefont {C.}~\bibnamefont
  {{B{\oe}hm}}}, \bibinfo {author} {\bibfnamefont {J.~A.}\ \bibnamefont
  {{Schewtschenko}}}, \bibinfo {author} {\bibfnamefont {R.~J.}\ \bibnamefont
  {{Wilkinson}}}, \bibinfo {author} {\bibfnamefont {C.~M.}\ \bibnamefont
  {{Baugh}}}, \ and\ \bibinfo {author} {\bibfnamefont {S.}~\bibnamefont
  {{Pascoli}}},\ }\Doi {10.1093/mnrasl/slu115} {\bibfield  {journal} {\bibinfo
  {journal} {\mnras},\ }\textbf {\bibinfo {volume} {445}},\ \bibinfo {pages}
  {L31} (\bibinfo {year} {2014})},\ \Eprint {http://arxiv.org/abs/1404.7012}
  {arXiv:1404.7012} \BibitemShut {NoStop}%
\bibitem [{\citenamefont {{Essig}}\ \emph
  {et~al.}(2012){\natexlab{b}}\citenamefont {{Essig}}, \citenamefont
  {{Mardon}},\ and\ \citenamefont {{Volansky}}}]{Essig_2012a}%
  \BibitemOpen
  \bibfield  {author} {\bibinfo {author} {\bibfnamefont {R.}~\bibnamefont
  {{Essig}}}, \bibinfo {author} {\bibfnamefont {J.}~\bibnamefont {{Mardon}}}, \
  and\ \bibinfo {author} {\bibfnamefont {T.}~\bibnamefont {{Volansky}}},\ }\Doi
  {10.1103/PhysRevD.85.076007} {\bibfield  {journal} {\bibinfo  {journal}
  {\prd},\ }\textbf {\bibinfo {volume} {85}},\ \bibinfo {eid} {076007}
  (\bibinfo {year} {2012}{\natexlab{b}})},\ \Eprint
  {http://arxiv.org/abs/1108.5383} {arXiv:1108.5383} \BibitemShut {NoStop}%
\bibitem [{\citenamefont {{Wilkinson}}\ \emph {et~al.}(2014)\citenamefont
  {{Wilkinson}}, \citenamefont {{Lesgourgues}},\ and\ \citenamefont
  {{B{\oe}hm}}}]{Wilkinson_2014}%
  \BibitemOpen
  \bibfield  {author} {\bibinfo {author} {\bibfnamefont {R.~J.}\ \bibnamefont
  {{Wilkinson}}}, \bibinfo {author} {\bibfnamefont {J.}~\bibnamefont
  {{Lesgourgues}}}, \ and\ \bibinfo {author} {\bibfnamefont {C.}~\bibnamefont
  {{B{\oe}hm}}},\ }\Doi {10.1088/1475-7516/2014/04/026} {\bibfield  {journal}
  {\bibinfo  {journal} {\jcap},\ }\textbf {\bibinfo {volume} {4}},\ \bibinfo
  {eid} {026} (\bibinfo {year} {2014})},\ \Eprint
  {http://arxiv.org/abs/1309.7588} {arXiv:1309.7588} \BibitemShut {NoStop}%
\bibitem [{\citenamefont {{Khatri}}\ and\ \citenamefont
  {{Sunyaev}}(2012)}]{Khatri2012mix}%
  \BibitemOpen
  \bibfield  {author} {\bibinfo {author} {\bibfnamefont {R.}~\bibnamefont
  {{Khatri}}}\ and\ \bibinfo {author} {\bibfnamefont {R.~A.}\ \bibnamefont
  {{Sunyaev}}},\ }\Doi {10.1088/1475-7516/2012/09/016} {\bibfield  {journal}
  {\bibinfo  {journal} {\jcap},\ }\textbf {\bibinfo {volume} {9}},\ \bibinfo
  {eid} {016} (\bibinfo {year} {2012})},\ \Eprint
  {http://arxiv.org/abs/1207.6654} {arXiv:1207.6654} \BibitemShut {NoStop}%
\bibitem [{\citenamefont {{Chluba}}\ and\ \citenamefont
  {{Jeong}}(2014)}]{Chluba2013PCA}%
  \BibitemOpen
  \bibfield  {author} {\bibinfo {author} {\bibfnamefont {J.}~\bibnamefont
  {{Chluba}}}\ and\ \bibinfo {author} {\bibfnamefont {D.}~\bibnamefont
  {{Jeong}}},\ }\Doi {10.1093/mnras/stt2327} {\bibfield  {journal} {\bibinfo
  {journal} {\mnras},\ }\textbf {\bibinfo {volume} {438}},\ \bibinfo {pages}
  {2065} (\bibinfo {year} {2014})},\ \Eprint {http://arxiv.org/abs/1306.5751}
  {arXiv:1306.5751} \BibitemShut {NoStop}%
\end{thebibliography}%

\end{document}